\documentclass[aps,prl,preprintnumbers,showpacs,twocolumn,groupedaddress]{revtex4}

\usepackage{amsmath,amssymb}
\usepackage{graphicx}
\usepackage{psfrag}

\newcommand{\lsbar}{\langle l_s\rangle}
\newcommand{\lmin}{l_{\rm min}}

\begin{document}

\preprint{LMU-ASC 23/06}

\title{Floppy modes and non-affine deformations in random fiber networks}
\author{Claus Heussinger} \author{Erwin Frey}

\affiliation{Arnold Sommerfeld Center for Theoretical Physics and
  Center for NanoScience, Department of Physics, Ludwig-Maximilians-Universit\"at
  M\"unchen, Theresienstrasse 37, D-80333 M\"unchen, Germany}

\begin{abstract}
  We study the elasticity of random fiber networks. Starting from a
  microscopic picture of the \emph{non-affine} deformation fields we calculate
  the macroscopic elastic moduli both in a scaling theory and a self-consistent
  effective medium theory. By relating non-affinity to the low-energy
  excitations of the network (``floppy-modes'') we achieve a detailed
  characterization of the non-affine deformations present in fibrous networks.

\end{abstract}

 \date{\today}

\maketitle

Materials as different as granular matter, colloidal suspensions or lithospheric
block systems share the common property that they may exist in a highly fragile
state~\cite{cates00,soloviev03}. While in principle able to withstand static
shear stresses, small changes in the loading conditions may lead to large scale
structural rearrangements or even to the complete fluidization of the
material~\cite{cates98,corwin05,soloviev03}. To understand the extraordinary
mechanical properties of these systems new concepts have to be developed that go
beyond the application of classical elasticity theory and that sufficiently
reflect the presence of the microstructure~\cite{alexander98}. One example is
the ``stress-only'' approach to the elasticity of granular
materials~\cite{wittmer96}, where the elimination of the kinematic degrees of
freedom accounts for the infinite stiffness of the grains. This seems to capture
the inhomogeneous distribution of stresses in the sample and their concentration
along the so called force-chains~\cite{bouchaud01}. In jammed systems of soft
spheres, on the other hand, fragility has recently been shown to directly
affect the deformation response of the system. While it may induce anomalous
deformation fields that strongly deviate from the expectations of homogeneous
elasticity (``non-affine'' deformations)~\cite{tanguy02}, it may also lead to a
proliferation of low-frequency vibrational states far beyond the usual
Debye-behaviour of ordinary solids~\cite{silbert05}. It has been argued that
these low-energy vibrations derive from a set of zero-frequency modes (floppy
modes) that are present just below the jamming threshold~\cite{wyart05b} and
relate to the ability of the structure to internally rearrange without any
change in its potential energy. This concept of floppy modes has also been used
in connection with elastic percolation networks where the fragile state is
reached by diluting a certain fraction of nearest-neighbour contacts. In these
systems constraint-counting arguments may be used to determine the percolation
transition at which the system looses its rigidity~\cite{jacobs95}.

Here our focus is on a particular class of heterogeneous networks composed of
crosslinked fibers. These systems have recently been suggested as model systems
for studying the mechanical properties of paper sheets~\cite{alava06} or
biological networks of semiflexible polymers~\cite{bausch06,heu06a}. While these
networks are known to have a rigidity percolation transition at low
densities~\cite{lat01,wil03}, we show here that even networks in the
high-density regime in many ways resemble the behaviour of fragile matter,
despite the fact that they are far away from the percolation threshold. We
identify the relevant floppy modes and highlight their importance for
understanding the macroscopic elasticity of the network. In particular, we will
be able to explain the occurence of an anomalous intermediate scaling regime
observed in recent simulations~\cite{wil03,hea03a,hea03c}.  In this regime the shear
modulus was found to depend on density (measured relative to the percolation
threshold) as $G\sim \delta\!\rho^\mu$ with a fractional exponent as large as
$\mu \approx 6.67$ \cite{wil03}. Also, highly non-affine
deformations~\cite{hea03a,onck05} as well as inhomogeneous distribution of
stresses in the network have been found. Heuristic non-affinity measures have
been devised~\cite{hea03a,onck05}, however, little is known about the actual nature
of the deformations present. While the expression ``non-affine'' is exclusively
used to signal the absence of conventional homogeneous elasticity, scarce
positive characterization of non-affine deformations has been achieved up to
now~\cite{didonna05}. This Letter tries to fill this gap by characterizing in
detail the non-affine deformation field present in fibrous networks. By relating
non-affinity to the floppy modes of the structure we can, starting from a
microscopic picture, calculate the macroscopic elastic moduli both in a scaling
theory and a self-consistent effective medium approximation. { In analogy with the
  affine theory of rubber elasticity for flexible polymer gels, our approach
  might very well serve as a second paradigm to understand the elasticity of
  microstructured materials.} Due to the proximity to the fragile state, it
might also be of relevance to force transmission in granular media and to the
phenomenon of jamming.

The two-dimensional fiber network we consider is defined by randomly placing $N$
elastic fibers of length $l_f$ on a plane of area $A=L^2$ such that both
position and orientation are uniformly distributed. We consider the fiber-fiber
intersections to be perfectly rigid, but freely rotatable cross-links that do
not allow for relative sliding of the filaments. The randomness entails a
distribution of angles $\theta\,{\rm\epsilon}\,[0,\pi]$ between two intersecting
filaments
\begin{equation}\label{eq:angleDist}
P(\theta) = \frac{\sin(\theta)}{2}\,,
\end{equation}
while distances between neighbouring intersections, the segment lengths $l_s$,
follow an exponential distribution~\cite{kal60}
\begin{equation}\label{eq:segDist}
  P(l_s)=\lsbar^{-1} e^{-l_s/\lsbar}\,.
\end{equation}
The mean segment length $\lsbar$ is inversely related to the line density
$\rho=Nl_f/A$ as $\lsbar= \pi / 2\rho $. The segments are modeled as classical
beams with cross-section radius $r$ and bending rigidity $\kappa$.  Loaded along
their axis (``stretching'') such slender rods have a rather high stiffness
$k_\parallel(l_s)=4\kappa/l_sr^2$, while they are much softer with respect to
transverse deformations $k_\perp(l_s)=3\kappa/l_s^3$ (``bending''). Numerical
simulations for the effective shear modulus $G$ of this network have identified
a cross-over scaling scenario characterized by a length scale
$\xi=l_f(\delta\!\rho l_f)^{-\nu}$ and $\nu\approx2.84$~\cite{wil03} that
mediates the transition between two drastically different elastic regimes. For
fiber radius $r\gg \xi$ the system is in an affine regime where the elastic
response is mainly dominated by stretching deformations homogeneously
distributed throughout the sample. The modulus in this regime is simply
proportional to the typical stretching stiffness, $G_{\rm aff}\propto
k_\parallel(\lsbar)$ and independent of the fiber length $l_f$.  This is in
marked contrast to the second regime at $r\ll \xi$. There, only non-affine
bending deformations are excited and the modulus shows a strong dependence on
fiber length $G_{\rm na} \propto k_\perp(\lsbar)(l_f/\lsbar)^{\mu-3}$. Using
renormalization-group language the parameters $r$ and $l_f$ may be viewed as
scaling fields (measured in units of the ``lattice-constant'' $\lsbar$). The
stretching dominated regime may then be characterized by an (affine) fixed-point
at $l_f\to\infty$ and finite radius $r\neq 0$.  On the other hand, the
(non-affine) fixed-point of the bending dominated regime is obtained by first
letting $r\to0$ and then performing $l_f\to\infty$. This suggests that the
elastic properties in the latter regime may be analysed at vanishing radius
$r=0$, that is by putting the system on the stable manifold of the fixed point.

In the following we will exploit this limit to calculate the modulus $G_{\rm
  na}$ in the non-affine regime. Central to the analysis is the recognition that
in this limit the ratio of bending to stretching stiffness
$k_\perp/k_\parallel\propto r^2$ tends to zero and bending deformations become
increasingly soft. We thus obtain the much simpler problem of a central-force
network.  However, as only two fibers may intersect at a cross-link the
coordination is $z<4$~\footnote{Due to the finite fiber length there are also
  two- and three-fold coordinated cross-links.}  and rigid regions may not
percolate through the system~\cite{kello96,maxwell1864}. This implies that on a
macroscopic level, the elastic moduli will be zero, while microscopically
displacements can be chosen such that segment lengths need not be changed. These
are the floppy modes of the structure that entail the fragility of the network
in the bending dominated regime. It has been argued that a critical coordination
of $z_c=4$ is necessary to give the network rigidity~\cite{maxwell1864}. This
value defines the ``isostatic'' point, which in our network corresponds to
taking the limit $l_f\to\infty$.  Thus, we arrive at the conclusion that
isostaticity and the onset of rigidity seem to be intimately connected to the
fixed point governing the non-affine regime. While it is usually not possible to
deduce the specific form of the floppy modes, the fibrous architecture allows
for their straightforward construction.
\begin{figure}[t]
 \begin{center}
  \includegraphics[width=0.9\columnwidth]{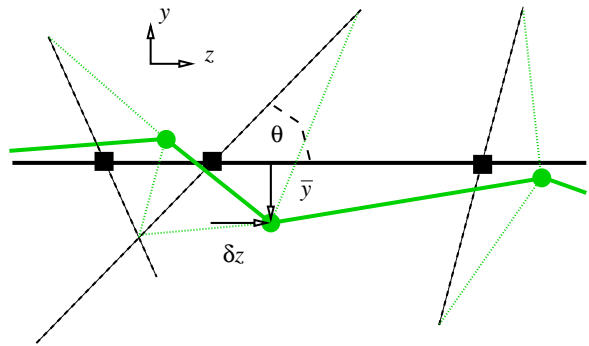}
\end{center}
\caption{Construction of a floppy mode by axial displacement $\delta z$ of the
  primary fiber (drawn horizontally) and subsequent transverse deflection $\bar
  y = -\delta z\cot\theta$ of the crosslinks to restore the segment lengths on
  the secondary fibers (dashed lines, possible to first order in $\delta z$).
  Initial cross-link positions are marked as black squares, final configurations
  as red circles.}
  \label{fig:floppyMode}
\end{figure}
In a first step (see Fig.\ref{fig:floppyMode}) we perform an arbitrary axial
displacement $\delta z$ of a given (primary) fiber as a whole. This, of course,
will also affect the crossing (secondary) fibers such that the lengths of
interconnecting segments change. In a second step, therefore, one has to account
for the length constraints on these segments by introducing cross-link
deflections $\bar{y}_i= -\delta z \cot\theta_i$ transverse to the contour of the
primary fiber. As a result all segment lengths remain unchanged to first order
in $\delta z$~\footnote{A similar construction holds in 3d, where, in addition,
  each cross-link acquires a floppy ``out-of-plane'' degree of freedom.}. The
construction is therefore suitable to describe the linear response properties of
the network, while at the same time it offers an explanation for the stiffening
behaviour found in fully nonlinear simulations~\cite{onck05,tbp}. {Any finite
  strain necessarily leads to the energetically more expensive stretching of
  bonds and therefore to an increase of the modulus.}

{The identified modes take the form of localized excitations that affect only
  single filaments and their immediate surroundings.} By superposition we may
therefore construct a displacement field that allows the calculation of
macroscopic quantities like the elastic moduli.  To achieve this we need to know
the typical magnitude of displacements $\delta z$ of a given fiber relative to
its surroundings, the crossing secondary fibers.  Since $\delta z$ is defined on
the scale of the complete fiber we do not expect any dependence on average
segment length $\lsbar$, such that $\delta z\propto l_f$ remains as the only
conceivable possibility. Alternatively, one may obtain the same result by
assuming that the individual fiber \emph{centers} follow the macroscopic strain
field in an affine way.  Then, relative displacements of centers of neighbouring
fibers would be proportional to their typical distance.  This is of the order
filament length $l_f$ and again $\delta z \propto l_f$.  Note, however, that the
assumption of affine displacement of the fiber centers cannot be literally true
for fibers intersecting at very small angles $\theta\to0$. To avoid a diverging
transverse deflection $\bar{y}_i=-\delta z \cot\theta\to\infty$ the two fibers
will most likely not experience any relative motion at all and $\delta z\to 0$.
Truly affine displacements can therefore only be established on scales larger
than the filament length. It should also be clear, that the assumption of affine
displacements of the fiber centers is different from the usual approach of
assigning affine deformations on the scale of the single segment.  The latter
would lead to deformations $\delta_{\rm aff} \propto l_s$, proportional to the
length $l_s$ of the segment. Instead, axial displacements of the fiber as a
whole are, by construction of the floppy mode, directly translated into {\it
  non-affine} deformations $\delta_{\rm na} \propto l_f$, which do not depend on
the length of the segment.

Restoring the radius $r$ to its finite value, the floppy modes acquire energy
and lead to bending of the fibers. A segment of length $l_s$ will then typically
store the energy $w_b(l_s)\simeq \kappa\delta_{\rm na}^2/l_s^3\simeq \kappa
l_f^2/\l_s^3$. By averaging over the segment length distribution
Eq.(\ref{eq:segDist}) one may calculate the average bending energy $\langle W_b
\rangle$, stored in a fiber consisting of $n\simeq\rho l_f$ segments,
\begin{equation}\label{eq:avgFiberEnergy}
  \langle W_b \rangle \simeq \rho l_f\int_{\lmin}^\infty dl_sP(l_s)
  \frac{\kappa\delta_{\rm na}^2}{l_s^3}\,.
\end{equation}
We assume the integral to be regularized by a lower cut-off length $\lmin$, that
we now determine in a self-consistent manner. Physically, $\lmin$ corresponds to
the shortest segments along the fiber that contribute to the elastic energy.
Even though we know (see Eq.(\ref{eq:segDist})) that arbitrarily short segments
do exist, their high bending stiffness $k_\perp(l_s)\propto l_s^{-3}$ makes
their deformation increasingly expensive. Segments with length $l_s < \lmin$
will therefore be able to relax from their floppy mode deformation $\delta_{\rm
  na}$, thereby reducing their bending energy from $w_b(\lmin)$ to nearly zero.
However, due to the length constraints this relaxation necessarily leads to the
movement of an entire secondary fiber and to the excitation of a floppy mode
there. By balancing $w_b(\lmin) = \langle W_b \rangle$ this gives $\lmin \simeq
1/\rho^2l_f$ and for the average bending energy of a single fiber $\langle
W_b\rangle \simeq \kappa/l_f(\rho l_f)^6$. This implies for the modulus $G_{\rm
  na}\simeq \rho/l_f\langle W_b\rangle\propto \rho^7$, which compares well with
the simulation result of $\mu = 6.67$. What is more, by equating the energy
$\langle W_b\rangle$ with $\langle W_s\rangle\simeq \kappa l_fr^{-2}$ valid in
the affine stretching regime, one can also infer the crossover exponent $\nu=3$.

In summary, we have succeeded in explaining the elasticity of the bending
dominated regime starting from the microscopic picture of the floppy modes that
characterize directly the deformation field deep inside the non-affine regime.
Alternatively, one might try to understand the emergent non-affinity in a
perturbative approach that considers deviations from an affine reference state.
Such a line of reasoning has recently been suggested in \cite{hea03c}, where
non-affine boundary layers, growing from the filament ends, are assumed to
perturb the perfect affine order.  However, comparing with their simulation data
the authors could not confirm the scaling picture unambiguously and acknowledged
the need for further numerical as well as improved theoretical
work~\cite{hea03c}. Thus, non-affine elasticity in fibrous networks appears to
be intrinsically a non-perturbative strong-coupling phenomenon for which the
floppy mode picture provides the correct low-energy excitations.  As we will
explicitly show next, one particular strength of our approach is that the
scaling picture can readily be extended to a full theory that self-consistently
calculates the modulus in a non-affine effective medium theory.

To set up the theory we consider a single filament together with its cross-links
that provide the coupling to the medium. The energy of this assembly consists of
two parts. First, the bending energy of the primary fiber
\begin{equation}\label{eq:}
  W_b[y(z)] = \frac{\kappa}{2}\int \left(\frac{\partial^2y}{\partial z^2}\right)^2dz\,,
\end{equation}
due to a transverse deflection $y(z)$. A second ``stretching'' energy
contribution arises whenever a cross-link deflection $y_i=y(z_i)$ differs from
its prescribed value $\bar y_i = -\delta z\cot\theta_i$ and may be written in
the form of an harmonic confining potential $W_s(y_i) = \frac{1}{2}k_i(y_i-\bar
y_i)^2$ that acts individually on each of the $n\simeq\rho l_f$ cross-links. It
allows the filament to reduce its own energy at the cost of deforming the
elastic matrix it is imbedded into. Performing a configurational average
$\langle . \rangle$ over cross-link positions $z_i$ and orientations $\theta_i$
we obtain the average elastic energy stored in a single fiber as
\begin{equation}\label{eq:selfconsEn}
  \langle W \rangle = \left\langle \min_{y(z)}\left( W_b[y(z)] + \sum_{i=1}^{n}
      \frac{k_i}{2}\left(y_i-\bar y_i\right)^2 \right)\right\rangle\,.
\end{equation}

To solve the model we further need to specify the stiffness $k_i=k(\theta_i)$ of
the medium that relates to the relaxation mode of a cross-link on the primary
filament from its floppy mode deflection. As we have argued above, any
relaxation of this kind must act as axial displacement on a secondary fiber,
thus exciting a new floppy mode there. The energy scale associated with this is
$\langle W\rangle$ such that we can write
\begin{equation}\label{eq:springConst}
  k(\theta_i)=2\langle  W\rangle\frac{\sin^2(\theta_i)}{\delta z^2}\,,
\end{equation}
where the angular dependence derives from the projection onto the axis of the
secondary filament. Eqs.(\ref{eq:selfconsEn}) and (\ref{eq:springConst})
represent a closed set of equations to calculate the configurationally averaged
deformation energy $\langle W\rangle$ as a function of the number of cross-links
$n$. In implementing this scheme we have generated ensembles of filaments with a
distribution of cross-linking angles as given by Eq.(\ref{eq:angleDist}) and
segment-lengths according to Eq.(\ref{eq:segDist}). Note, that there is no free
parameter in this calculation.
\begin{figure}[t]
 \begin{center}
  \includegraphics[width=0.9\columnwidth]{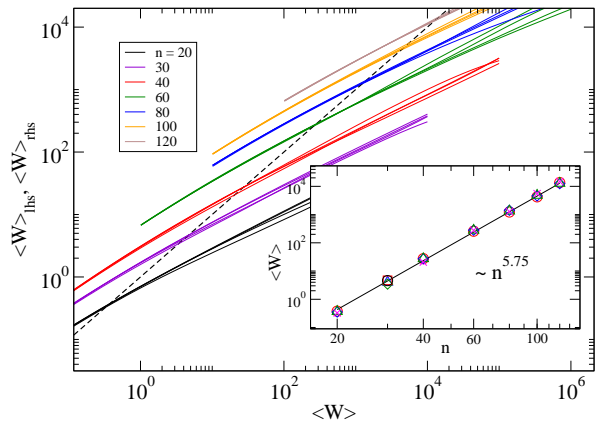}
\end{center}
\caption{ Graphical solution of Eqs.(\ref{eq:selfconsEn}) and
  (\ref{eq:springConst}) for various numbers $n$ of cross-links obtained by
  calculating the intersection between the left side of the equation $\langle
  W\rangle_{\rm lhs}$ (bisecting line, dashed curve) with the right side
  $\langle W\rangle_{\rm rhs}$ (full curves). The different curves for a given
  $n$ correspond to ensembles of varying size. They seem to diverge in the limit
  $\langle W \rangle_{\rm rhs}\gg \langle W \rangle_{\rm lhs}$. In fact, there (and only
  there) the averaging procedure is ill defined~\cite{tbp}. Inset: Resulting
  dependence of $\langle W\rangle$ on $n$.}
  \label{fig:enScaling}
\end{figure}
The equations are solved graphically in Fig.\ref{fig:enScaling} by plotting both
sides of Eq.(\ref{eq:selfconsEn}) as a function of $\langle W \rangle$. The
point of intersection, which solves the equation, is shown in the inset as a
function of the number of cross-links $n$. For the same parameter window as used
in the network simulations~\cite{wil03}, it yields the scaling behaviour of
$\langle W\rangle\propto n^{5.75}$. This implies for the modulus the exponent of
$\mu = 6.75$, which improves upon the simple scaling picture presented above and
provides a very accurate calculation of the scaling exponent $\mu$.

In conclusion, we have succeeded in deriving the macroscopic elasticity of
random fibrous networks starting from a microscopic description of the
displacement field in a manner that does not rely on the notion of affine
deformations. We have given a floppy mode construction that may be applied to
any two or three-dimensional network with fibrous architecture, for example
paper or biological networks of semiflexible filaments. It may also be shown to
be relevant to systems where the constraint of straight fibers is
relaxed~\cite{tbp}. The unusually strong density dependence of the modulus found
here is a consequence of the exponential segment length distribution
(\ref{eq:segDist}) and the presence of the length-scale $l_{\rm min}$. While
identification of the floppy modes has been recognized to be highly important
for a description of force transmission in granular media or the jamming
transition in colloidal systems, one can rarely give the exact form of these
zero-energy excitations. On the contrary, we have achieved an explicit
construction of the floppy modes that can be put in the form of localized
elementary excitations (``floppions'') affecting only single filaments and their
immediate surroundings.

\begin{acknowledgments}
It is a pleasure to acknowledge fruitful discussions with David Nelson and Mikko
Alava.
\end{acknowledgments}


\end{document}